\documentclass{article}
\usepackage{times}
\usepackage{soul}
\usepackage{url}
\usepackage[hidelinks]{hyperref}
\usepackage[utf8]{inputenc}
\usepackage[small]{caption}
\usepackage{graphicx}
\usepackage{amsmath}
\usepackage{amsthm}
\usepackage{booktabs}
\usepackage{algorithm}
\usepackage{algpseudocode}
\usepackage{caption}
\usepackage{subcaption}
\usepackage{color}
\usepackage[british,UKenglish]{babel}
\usepackage{xcolor}
\begin{document}
\begin{center}
    \begin{Large}A Cross-Frequency Protective Emblem:\\Protective Options for Medical Units and Wounded Soldiers\\in the Context of (fully) Autonomous Warfare
    \end{Large}
\end{center}

\begin{center}
Daniel C. Hinck$^2$, Jonas J. Schöttler$^1$, Maria Krantz$^{1}$, \\Katharina-Sophie Isleif$^{1}$, Oliver Niggemann$^{1}$
\end{center}
$\ $
\begin{center}
  \begin{minipage}{0.45\textwidth} 
  \begin{center}
  $^1$ Institute for Automation Technology\\
  Helmut Schmidt University\\
  University of the Federal Armed Forces Hamburg\\
  Holstenhofweg 85, 22043 Hamburg\\
  \end{center}
  \end{minipage}
  \hfill
  \begin{minipage}{0.45\textwidth}
  \begin{center}
  $^2$ Faculty of Medical Service and Health Sciences\\
  Command and Staff College\\German Federal Armed Forces\\
  Manteuffelstr. 20, 22587 Hamburg\\
  \end{center}
\end{minipage}
\end{center}
$\ $\\$\ $
\begin{center}
    protective$\_$sign@hsu-hh.de
\end{center}

\subsection*{Abstract} 

\textit{The protection of non-combatants in times of (fully) autonomous warfare raises the question of the timeliness of the international protective emblem. Incidents in the recent past indicate that it is becoming necessary to transfer the protective emblem to other dimensions of transmission and representation. (Fully) Autonomous weapon systems are often launched from a great distance to the aiming point and there may be no possibility for the operators to notice protective emblems at the point of impact. In this case, the weapon system would have to detect such protective emblems and, if necessary, disintegrate autonomously or request an abort via human-in-the-loop. In our paper, we suggest ways in which a cross-frequency protective emblem can be designed. On the one hand, the technical deployment, e.g. in the form of RADAR beacons, is considered, as well as the interpretation by methods of machine learning. With regard to the technical deployment, possibilities are considered to address different sensors and to send signals out as resiliently as possible. When considering different signals, approaches are considered as to how software can recognise the protective emblems under the influence of various boundary conditions and react to them accordingly. In particular, a distinction is made here between the recognition of actively emitted signals and passive protective signals, e.g. the recognition of wounded or surrendering persons via drone-based electro-optical and thermal cameras. Finally, methods of distribution are considered, including encryption and authentication of the received signal, and ethical aspects of possible misuse are examined.}
\\\\
\textbf{Keywords:} Protective Emblem $\cdot$ Technical protective Signal Processing $\cdot$ Artificial Intelligence $\cdot$ Fully autonomous warfare $\cdot$ Pose estimation

\section{Introduction} 
\label{sec:Intro}

The Red Cross was first recognised as an internationally accepted distinctive emblem for the protection of wounded military personnel in armed conflicts in the First Geneva Conventions of 1864, titled "Übereinkunft zur Linderung des Looses der im Felddienste verwundeten Militärs". 
The precise depiction of this international emblem was clarified in Article 18 of the "Protocol Additional to the Geneva Conventions of 12 August 1949, and Relating to the Protection of Victims of International Armed Conflicts (Protocol I)". In summary, it involves the identification through a flag/surface that carries one of the four distinctive emblems, which may be visible with infrared devices. In addition to this passive representation, active light or radio signals, or electronic markings such as radar beacons, can be used to identify a protected facility or means of transportation. As a notable "digital" marking, the transmission of Global Positioning System (GPS) data from protected facilities to parties involved in the conflict is worth mentioning. While efforts have been made in recent decades to develop autonomous and fully autonomous weapon systems for more precise and effective operations, the visibility of the protective emblem has not kept pace with technological advancements.

After attacks on hospitals as part of "critical infrastructure" (Kritische Infrastruktur, KRITIS) in the cyberspace, discussions have arisen regarding the possibilities of protecting designated facilities (e.g. protected domains in cyberspace). However, the "visible" protective emblem and its protection have been overlooked, despite instances in recent history where designated facilities have been directly targeted for attacks, even with the establishment of protected zones through transmitted GPS data (e.g., Kunduz, Afghanistan, 2015; al-Atareb, Syria, 2021). In addition to these two events, the use of grenades dropped by a man-in-the-loop drone on an apparently injured soldier in the Ukrainian-Russian conflict has led to an imperative for the representation and respect of the protective emblem in the electromagnetic spectrum. 
The following article focuses on the most commonly used sensors and their modes of operation in autonomous weapon systems and derives a possible recognisable depiction of the protective distinctive emblem by these sensors. Another focus of this article is the potential perception of protective emblems and non-combatants by fully autonomous systems (drones).

\section{Technical background, needs and possible applications} 
\label{sec:TBNPA}
The technical aspects of (fully) autonomous weapon systems.

Long-range weapons are military projectiles or rockets that are launched from various delivery systems such as aircraft, watercraft, land vehicles, or handheld tube weapons systems (e.g. RGW 90 LRMP) at an unspecified distance from the target. The projectile or rocket is guided remotely or self-guided to the target.
Large-calibre weapons (e.g. Panzerhaubitze 2000) and rocket weapons (e.g. Multi Launch Rocket System, MARS 2) are classified under the term artillery and are typically land- or water-based weapon systems.

Modern fully autonomous long-range or artillery weapon systems have a higher effectiveness in terms of impact and accuracy compared to "conventional" weapon systems. As a result, these systems are designed to minimise or ideally prevent collateral damage to civilian infrastructure and non-combatants.

In the 1980s, particularly the U.S. military recognised the advantage of Precision Guided Munitions (PGM), specifically in the form of smart munitions, for artillery systems. This type of ammunition has the ability to autonomously search for, identify, and attack targets. The warheads are ejected as submunitions from a projectile casing, and a multitude of submunitions engage multiple targets in a defined area using autonomous target-seeking capabilities (such as proximity fused or guided munitions) \cite{mangus1987smart}. In the 1990s, this type of munition was referred to as artificial intelligence-based or autonomous munitions \cite{zahnd1995Kampf}.
During the Gulf War (1990-1991), approximately 90,000 tons of bombs were dropped by U.S. aircraft on Iraq and Kuwait. 7
The decision cycle for target engagement with the aforementioned weapon systems remains still under human control, meaning that at a defined point in time, a human consciously decides to engage a target (see loitering weapons). This decision cycle is divided into the decision points find, fix, track, target, engage, and assess (F2T2EA).
In contrast, fully autonomous weapon systems (Lethal Autonomous Weapon Systems, LAWS) undergo the above mentioned decision cycle for target engagement without further human control/decision-making after their activation.

In consequence warfare is increasingly conducted at a distance using stand-off weapons without supplementary human visual on-site target designation or visual target verification, either fully autonomously or semi-autonomously. As a consequence, this can result in the (un)intentional engagement of facilities marked only by analog-visible protective emblems such as red cross flags or color-coded markings on buildings. Particularly in the case of fully autonomous weapon systems, the question arises as to how the protective emblem should be evaluated in the F2T2EA decision cycle.

\section{Active protection emblem}
For the detection of active protective emblems, we can make a distinction between active detection and passive recognition. Active detection refers to the active emission of a signal by the protected entity, while passive recognition means the attacking system needs to recognise a protective emblem like a red cross or a surrendering soldier. 
 For example, in recent studies, automatic traffic sign recognition systems were already explored as a means of improving road safety. In these systems, cameras are mounted on the vehicle to capture video feeds of the road and recognise traffic signs, providing the driver with timely warnings, nowadays even in teal-time based on an embedded platform that employs digital image processing algorithms \cite{Bilgin_2016,Sridevi_2017}. 
However, traditional feature extraction techniques can be time-consuming and complex. Convolutional neural networks have shown promising results in improving the efficiency and robustness of these techniques \cite{Jung_2016,Hatolkar_2018}.  In the following, we will discuss possible technologies to achieve active detection and recognition. 

\subsection{Signal-based protective options} 
\label{sec:TSPO}

To create the cross-frequency protective emblem, we propose using a combination of low-frequency and high-frequency signals in a specific pattern. Autonomous weapon systems are often launched from long distances to their intended targets, making it difficult for operators to notice protective emblems at the point of impact. Therefore, it is important for these weapon systems to detect protective emblems and autonomously disintegrate or request an abort through human intervention. This paper proposes a cross-frequency protective emblem design that can be deployed through RADAR beacons or other means to target different sensors and send out signals that are resilient. The values for different signals listed in Table~\ref{table:signals} are general estimates and should be considered as approximate ranges. Note that the communication range is highly dependent on various factors such as atmospheric conditions, antenna gain, and transmitter power, and can vary significantly in practice. 
\
\begin{table*}[h!]
\centering
\begin{tabular}{|p{2cm}|p{1.5cm}|p{2cm}|p{3cm}|p{3cm}|}
\hline
Technology & Signal Frequency & Range  & Advantages & Disadvantages  \\ \hline
L-band$^*$ & 1-2 GHz & $\approx 1000\,\mathrm{km}$ & Good penetration, low attenuation, suitable for long-range communication and sensing & Lower data rates compared to higher frequency bands  \\ \hline
X-band$^*$ & 8-12 GHz & $\approx 100\,\mathrm{km}$ & High data rates, good for long-range communication and sensing, better resistance to interference & Limited penetration of solid objects such as buildings or trees, can be affected by atmospheric conditions  \\ \hline
Microwave$^*$ & 1-100 GHz & $\approx 100\,\mathrm{km}$ & Good for long-range communication and sensing, can penetrate some solid objects & Limited penetration of dense objects such as walls or buildings  \\ \hline
Infrared & 1-100 THz & $\approx 100\,\mathrm m$ &  Can detect heat signatures, good for night vision, passive detection & Limited range, susceptible to interference from other heat sources   \\ \hline
Optical & 430-750 THz & $\approx$ kilometers &  High resolution, good for visual imaging, active or passive sensing & Limited range in adverse weather, requires line-of-sight  \\ \hline
Thermal Camera & 9-14 µm & $\approx$ kilometers &  Can detect heat signatures, good for night vision, can penetrate some materials & Limited resolution, susceptible to interference from other heat sources  \\ \hline
RFID & kHz-GHz & $\approx$ meters \cite{Costa_2021} \newline $\approx 2\,\mathrm{km}$ \cite{Amato_2015} &  Low cost, small form factor, contactless, suitable for inventory tracking and asset management & Limited range, susceptibility to interference from other RFID devices \\ \hline
WiFi & 2.4 GHz, 5 GHz & $\approx 100\,\mathrm m$ &  High data rates, widespread availability, suitable for local area network (LAN) communication & Limited range, susceptible to interference from physical obstructions and other WiFi devices \\ \hline
Passive Signs & N/A & $\approx 100\,\mathrm m$ &  Can be used for detection and recognition of predefined signs or symbols & Limited to predefined signs, susceptible to false positives or negatives  \\ \hline
Night Vision Devices & N/A & $\approx 100\,\mathrm m$ & Can amplify low levels of light, good for night vision & Limited range, susceptible to interference from bright light sources and atmosphere \\  \hline
\end{tabular}
\caption{Signals and technologies for active and passive communication and recognition. $^*$ communication via satellite possible, bands used for RADAR }
\label{table:signals}
\end{table*}

In addition to the choice of technology and frequency band, signal distribution is an important consideration for communication and sensing systems. Bi-directional signal distribution allows for two-way communication, with signals being transmitted and received between the sender and receiver. This can be further classified into passive and active signals. Passive signals, such as those emitted by wounded or surrendering soldiers, can be detected and recognised by sensors such as thermal or optical cameras as discussed in Section~\ref{subsec:PR}. Active signals, on the other hand, are signals intentionally transmitted by the sender, such as RADAR beacons or RFID tags.

Signal distribution can also involve encryption and authentication of the received signal, which can help prevent unauthorised access or interference. Encryption involves encoding the signal in a way that can only be deciphered by authorised parties with the appropriate key or password as realised using PRN codes for identifying satellites. Authentication involves verifying the identity of the sender and receiver to ensure that the signal is being transmitted and received by the intended parties.

\begin{figure}[h]
  \centering
  \includegraphics[width=0.49\textwidth]{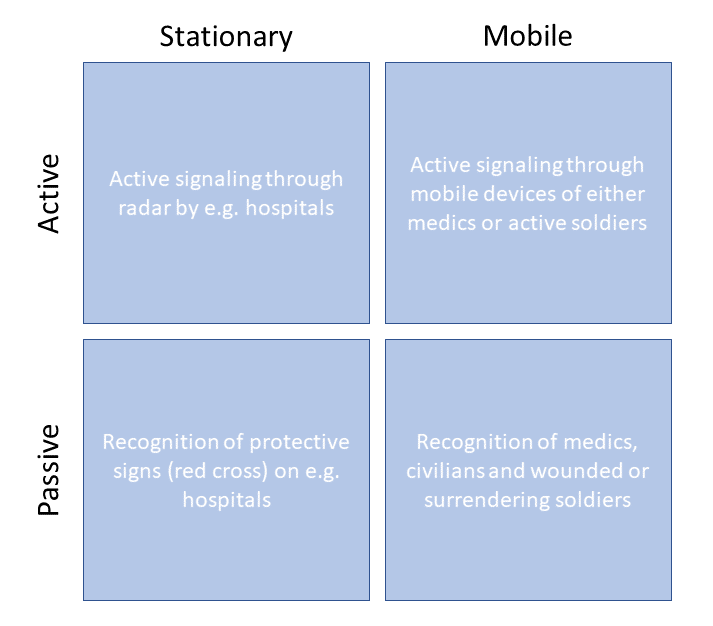} 
  \caption{The recognition of a protective emblem can be divided into active and passive recognition. Active recognition here refers to the active sending of a signal which marks an entity as being protected, e.g. by sending out a radar signal. Passive protective means a right for protective needs to be recognised by the attacking system, e.g. the red cross flag or wounded soldiers. Furthermore, both stationary and mobile entities can be protected.}
  \label{fig:matrix}
\end{figure}

\subsection{Software-based detection of active protective signs} 
\label{subsec:AR}

Active detection is always possible, when the protected entity actively emits a signal, for example using Radar, RFID or GPS. This is possible for stationary facilities (hospitals, civilian buildings), but could also be installed for tents used to tend wounded soldiers. Furthermore, active detection of mobile troops would be possible, as will be discussed below. 

\subsubsection{Detection of Stationary Facilities}
\label{subsec:DSF}

Stationary facilities are the easiest to protect when it comes to the emission of active signals, since emitters can be installed stationary. To ensure a safe detection of these facilities, a fusion of different signals would be the best option. Therefore, the facility would have to be equipped with several emitters. 

\textbf{RFID}, or Radio Frequency Identification \cite{xiao2007rfid}, uses radio waves to identify and track objects. The technology is already used to control access to important military areas, such as control centres or weapon systems. It consists of three basic components: a reader, an antenna and a tag.
The reader, which wants to identify an objects, emits radio waves in a certain frequency range, typically 125kHz, 13.56MHz, or 900MHz. 
The tag, which is positioned on the object to be tagged, contains the antenna and a microchip. On this microchip, a unique identification number and other data can be stored. Once a signal from the reader is captured by the tag, it sends back the information on the chip to the reader on the same frequency, which can process it using a computer system. 
While passive RFID tags don't have a power source and need the energy from the reader's signal to transmit their data, active RFID systems contain an energy source and can therefore actively transmit data. 
Active RFID systems also have a longer range and can transmit their signal over a range of a few meters to hundreds of meters, or even several kilometres. Active RFID systems can also be used for real-time location tracking. 
However, the range of an RFID signal is heavily dependent on its environment. Radio signals in the surrounding area, like cellular or Wi-Fi signals, can interfere with the RFID signal. Furthermore, physical barriers like walls can also block the signal. 
Once a signal is received by the reader, it first converts it from analogue to digital using a analogue-to-digital converter (ADC), before extracting the information contained in the signal. 
Since the signal could be distorted by interference, error correction algorithms are used. Thereby, RFID systems can improve the reliability and accuracy of the data being transmitted. Possible algorithms are Cyclic Redundancy Check (CRC), which adds a checksum to the transmitted data, or Forward Error Correction (FEC), which adds redundant information. 

\textbf{RADAR} is a technology that uses radio waves to detect the presence, location, and velocity of objects. It works by emitting a radio wave signal from a transmitter, which travels through the air and reflects off an object in its path. The reflected signal, or "echo," is then detected by a receiver and analysed to determine the distance, angle, and speed of the object.

In terms of detection, there are several factors that can affect a radar system's ability to detect objects. These include the range of the radar system (i.e. the maximum distance at which it can detect objects), the power and frequency of the radar signal, and the size and reflectivity of the object being detected. Additionally, factors such as atmospheric conditions, interference from other sources, and the presence of clutter (i.e. other objects in the radar's field of view) can also affect detection performance.

A radar signal can also be emitted by a stationary sender to transmit information to the receiver. This would be the use envisioned here. A sender installed in the protected facility would emit a radar signal, which can be received by the attacking weapon system. When a weapon system receives a radar signal, it typically uses the information provided by the radar to determine the location of the emitter. 

Since many facilities are equipped with radar emitters, many radar signals will be received by the weapon system simultaneously. It would therefore be important to ensure that the signals emitted by a protected facility are given priority. This could be hard-coded into the code e.g. once a signal associated with a protected facility is received it overrides all others. A better detection and discrimination of radar signals could also be achieved through the use of AI [Jonas Paper]. 

\textbf{GPS} (Global Positioning System), is a satellite-based navigation system that provides location and time information anywhere on or near the Earth. The GPS system consists of a network of satellites orbiting the Earth, as well as ground control stations and GPS receivers. Each satellite transmits a signal that contains information about its location and time.

To be able to use GPS, a weapon system must be able to receive signals from at least four of the GPS satellites. It then compares the time stamps of the signals received from the satellites and uses the differences in the time stamps to calculate the distance between itself and the satellites. Through trilateration the system can then determine its own position. 

To ensure safe detection of a protected facility, several signals should be combined. This would, for example, be possible by combining radar, GPS and RFID. A protected facility could send out a radar signal. Once this is received by a weapon system, it could determine the position of the emitting facility by using GPS to determine its own position and the radar signal it received from the facility. This information it could use to send a request to a database, to check whether a protected facility has been assigned this position. Once this is checked, the weapon system could send its own signal to the facility, based on RFID technology, and wait for the signal to be send back by the facility's RFID tag. Once it has received this signal, it would have to stop its attack and safely manoeuvre out of the protected zone. 

A problem arises when one of these signals is not available, for example when it is not possible to send a request to a database. It could be possible to also add a human-in-the-loop, who has to interfere in unclear cases. Furthermore, it would also be possible to always require a human to be involved in the decision of aborting an attack when a protective emblem is received. 

This approach would necessitate the installation of several sensors on a weapon system, as well as the ability to safely communicate with a database. For new weapon systems, these features could be demanded by international standards. Older systems would have to be retro-fitted, or in case that that is not possible, disused. 

\subsubsection{Detection of Mobile Units}
\label{subsec:DOT}

A problem arises when detecting moving protected entities, like wounded or surrendering soldiers or mobile paramedic units. These cannot be equipped with a radar signal and would therefore not be detected by the system described in the previous section.

A possible option would be to equip all weapons with RFID tags which could be read by an attacking weapon system. Units not carrying an RFID tag would be deemed protected and would not be attacked. 
RFID technology is already used to keep track of military equipment. RFID tags can be attached to vehicles, weapons, and other equipment to track their location and status. This can help military commanders to quickly locate and deploy assets, as well as monitor the maintenance and repair status of equipment \cite{buckner2002miclog}. RFID tags could also be attached to personnel to track their location and movements, as well as monitor their health and well-being \cite{nicholls2017implanting}.

A problem with detection of the tags arises when whole units are equipped with RFID tags. Especially when multiple tags are present within the range of a single reader, anti-collision protocols become necessary. Without an anti-collision protocol, the reader might not be able to distinguish between multiple tags and might read them all simultaneously or fail to read any of them.
Several different anti-collision protocols can be used with RFID systems, for example ALOHA protocol, tree based protocols, binary search algorithms or bitwise arbitration algorithms \cite{klair2010survey}.
The specific protocol used will depend on the application and the requirements of the system. The goal of these protocols is to ensure that each tag is identified in a timely and efficient way, while minimising collisions and other sources of interference.
ML approaches have been suggested to improve collision-free reading of RFID tags \cite{mafamane2022dmlar}. The paper proposes an anti-collision protocol called DMLAR that uses feed-forward Artificial Neural Network methodology to predict collisions and ensure efficient resource allocation in RFID networks. Such an approach might be necessary when a multitude of tags need to be read to identify armed military personnel. 

In situations where many RFID tags need to be read simultaneously, ML algorithms can also be applied to process and analyse the data. 
ML approaches have been applied to a variety of RFID data \cite{mohsen2022machine} \cite {tao2021remote} \cite{senta2007machine}. 
It is for example possible to use data collected from RFID tags to train ML algorithms for anomaly detection. A study aimed to develop a system that detects abnormal behaviour in elderly people at home using active RFID tags \cite{hsu2010rfid}. Movement data was collected through the RFID reader's signals, clustering techniques were used to build a personalised model of normal behaviour, and any incoming data outside the model is viewed as abnormal and triggers an alarm. Similarly, algorithms could be trained to recognise the movement patterns of active soldiers as compared to injured soldiers, based on the data received from RFID tags. 

The feasibility of this approach of recognising protected individuals (or, better, not protected individuals) depends on the willingness of all nations to participate in such an approach. Two major points need to be addressed: 
\begin{enumerate}
    \item All weapon manufactures which produce hand-held, portable weapon systems would have to equip their weapons with RFID tags, which work on an internationally assigned frequency.  
    \item All existing hand-held, portable weapons would have to be retro-fitted with RFID tags operating on the internationally assigned frequency. 
\end{enumerate}

\section{Passive protective emblem} 

\label{subsec:PR}
According to the Geneva Convention, every non-combatant is to be protected from any act of violence and war. In most cases, those do not wear or carry any protective emblem and if they do, for example medic units, it might be a flag or a symbol worn on the persons. But mostly they do not and therefore they are to be identified as non-combatants in a different way, for example on the perception of their status as a surrendering, wounded or unarmed person. A (fully) autonomous weapon system has to recognise these non-combatants by itself and if possible, inform the operator about the recognition. 
\\\\
The protective emblem here includes the condition, situation and status of a person. This comprises injuries, medical supplies and resulting persons. These situation-based protective emblems must be determined and evaluated on the basis of image and video material recorded by electro-optical (EO) or infrared (IR) sensors, from drones for example. Figure \ref{fig:drone_images} shows some examples just before the dropping of payload on (wounded) solders from drones in recent conflicts. A distinction must be made between persons to be protected, who have an appropriate status, and soldiers in lurking positions or ambushes. The second do not fall under the protection of the Geneva Convention. 
\\\\
The current use of the protective emblem does not, in respect to new development in weapons technology, cover all those conditions mention above. The protective emblem therefore has to be enhanced and developed further to ensure that the protection of non-combatants is still (as best as possible) guaranteed. Therefore, the protective emblem could make use of modern sensor technologies and take the step from the marking of the to-be-protected to the recognition of those, hence extend the responsibility from the non-combatants to the offenders and equip systems with appropriate routines. In the following sections we will describe how this could be done for (fully) autonomous systems using sensors to take notice of passive protective emblems. 

\subsubsection{Definition of passive protective emblems}
\label{sec:def_pass_sign}
A passive protective emblem is, according to our distribution, every sign that is not transmitted. Passive protective emblems could also include the active action of persons, for example when deploying flags or signs with the classic protective emblem. However, these are not send out as active transmissions. 
\\\\
Furthermore, a protective emblem can be derived from the situation in which a person finds him/herself. These situations are every possible situation except taking part actively in fighting and combat. Based on that, the detection of this kind of emblems is not the recognition of non-combats, but rather the recognition of combatants and the excluding of every other person "on the field". 
\subsubsection{EO/IR-based recognition of protective emblems with AI}
\label{subsec:EO_IR_RPS}
\begin{figure*}[h]
  \centering
  \includegraphics[width=0.49\textwidth]{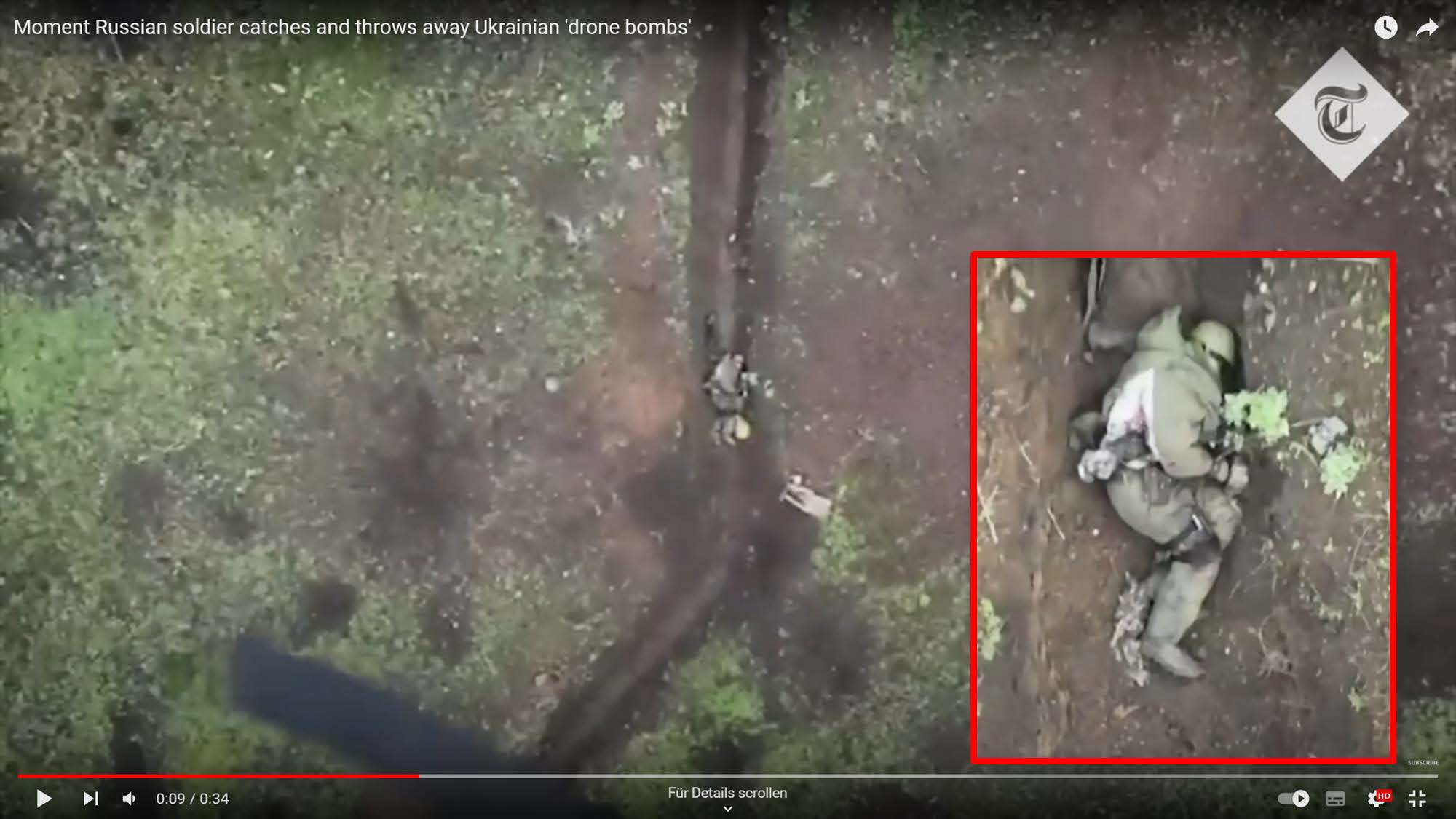}
  \hfill
  \includegraphics[width=0.49\textwidth]{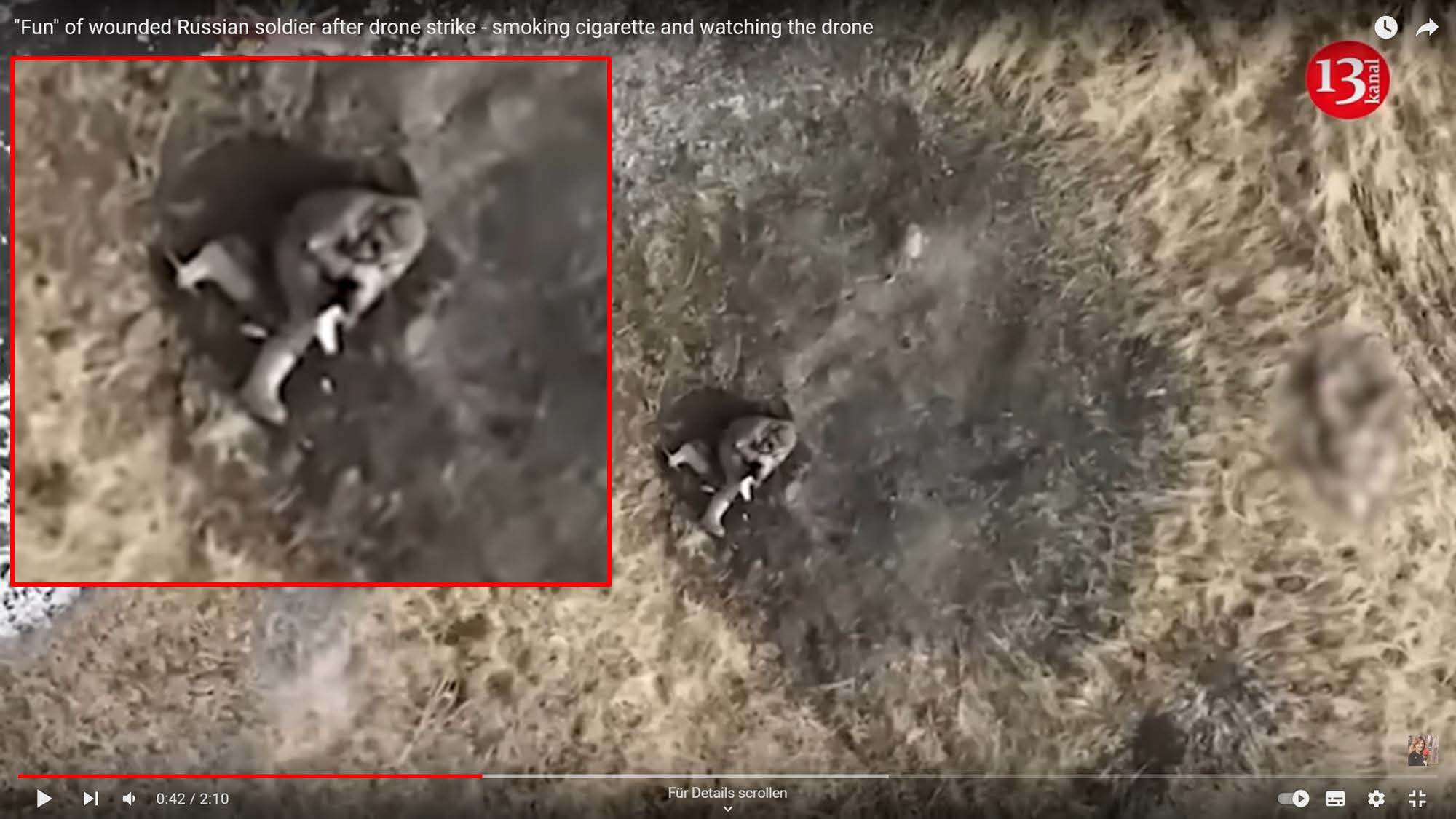}
  \vfill
  \includegraphics[width=0.49\textwidth]{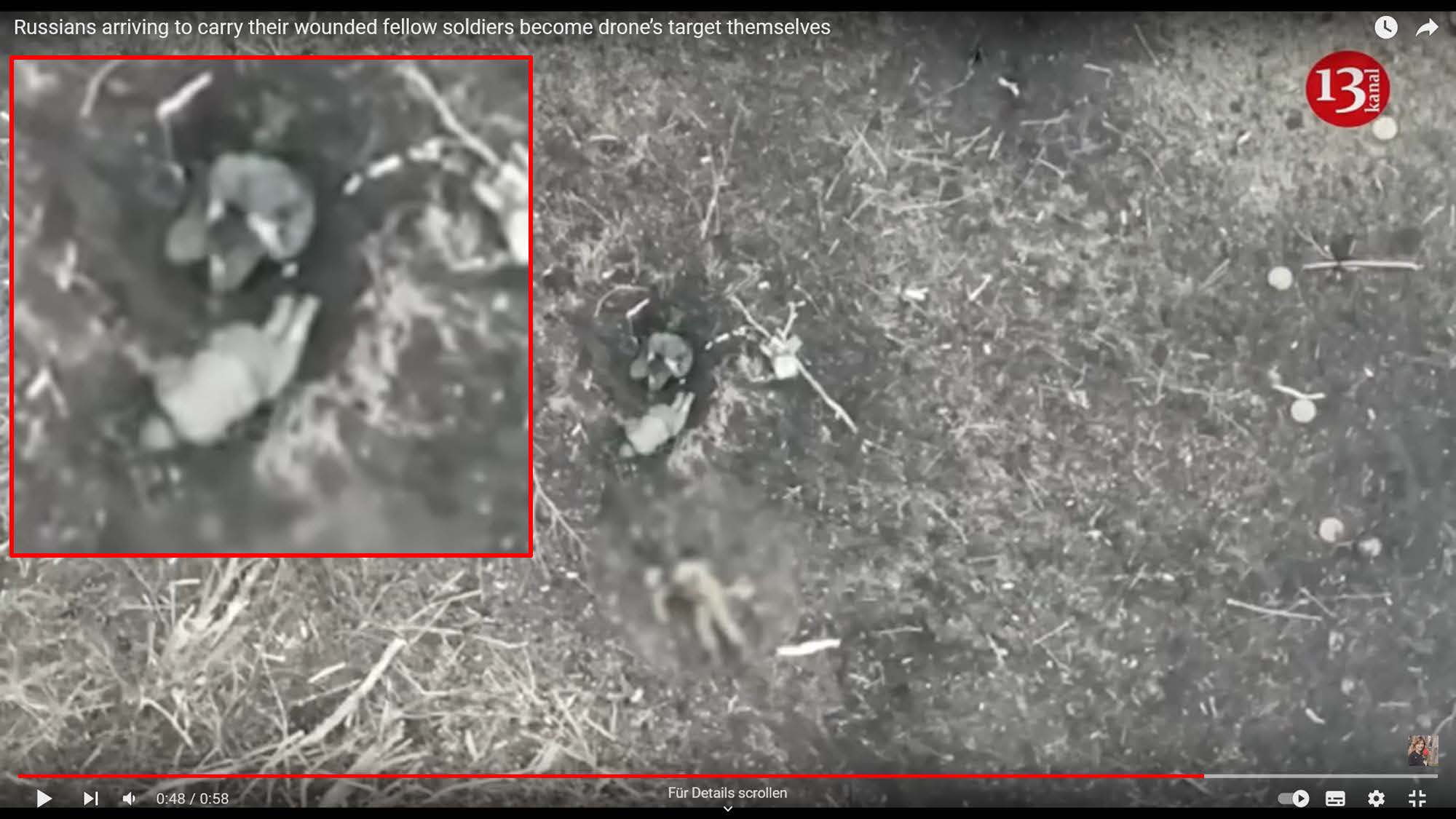}
  \hfill
  \includegraphics[width=0.49\textwidth]{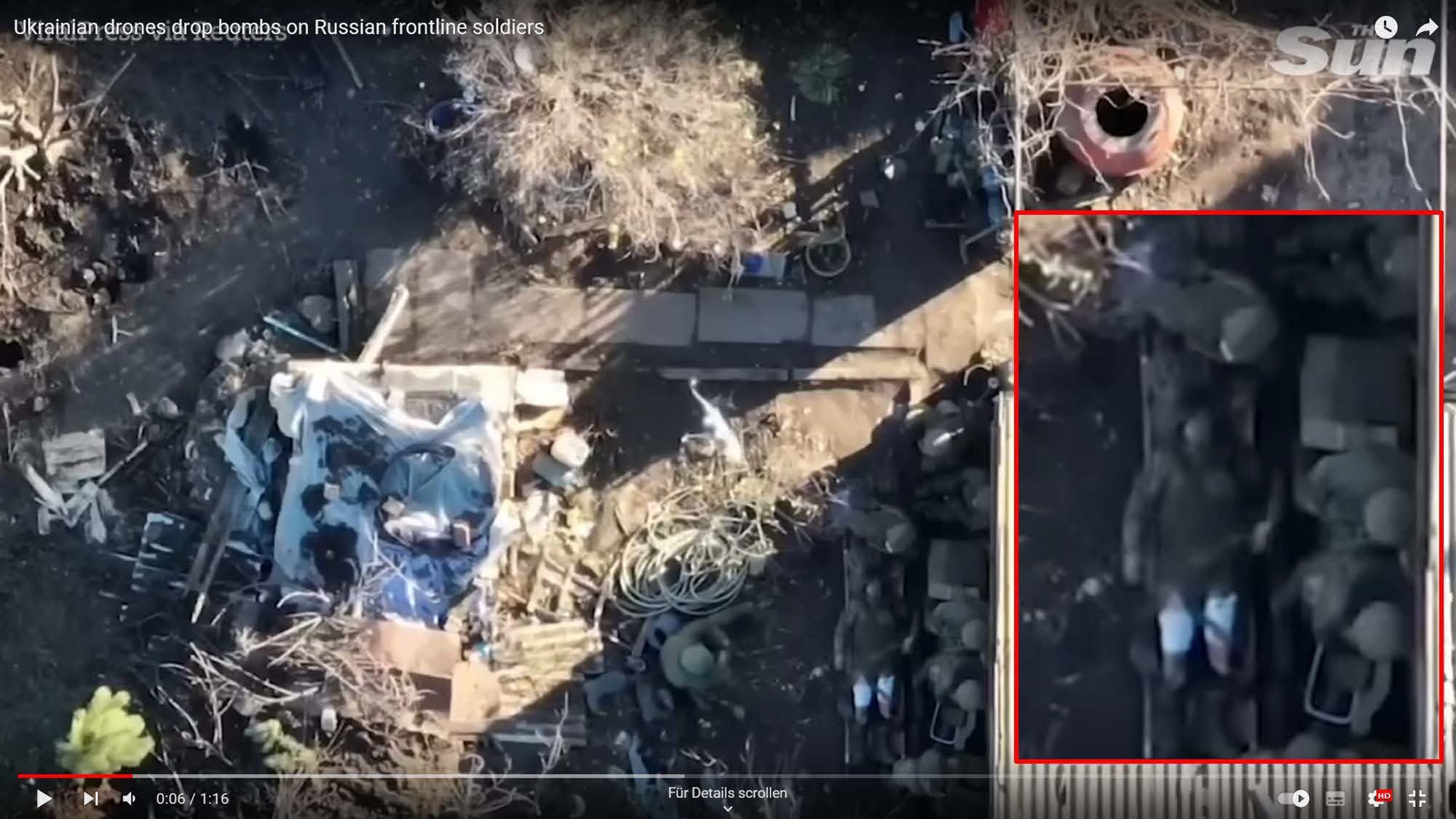}
  \vfill
  \includegraphics[width=0.49\textwidth]{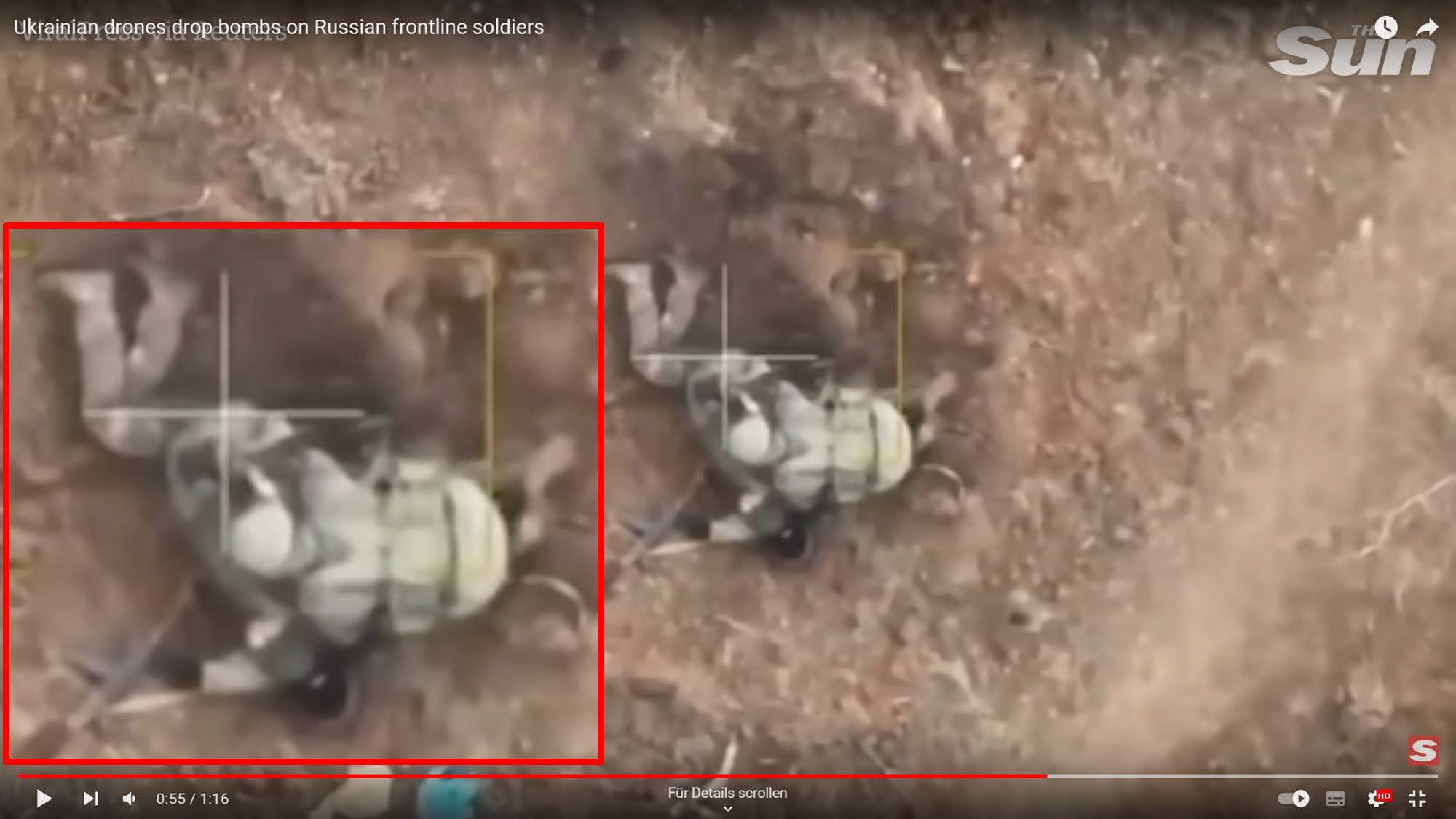}
  \hfill
  \includegraphics[width=0.49\textwidth]{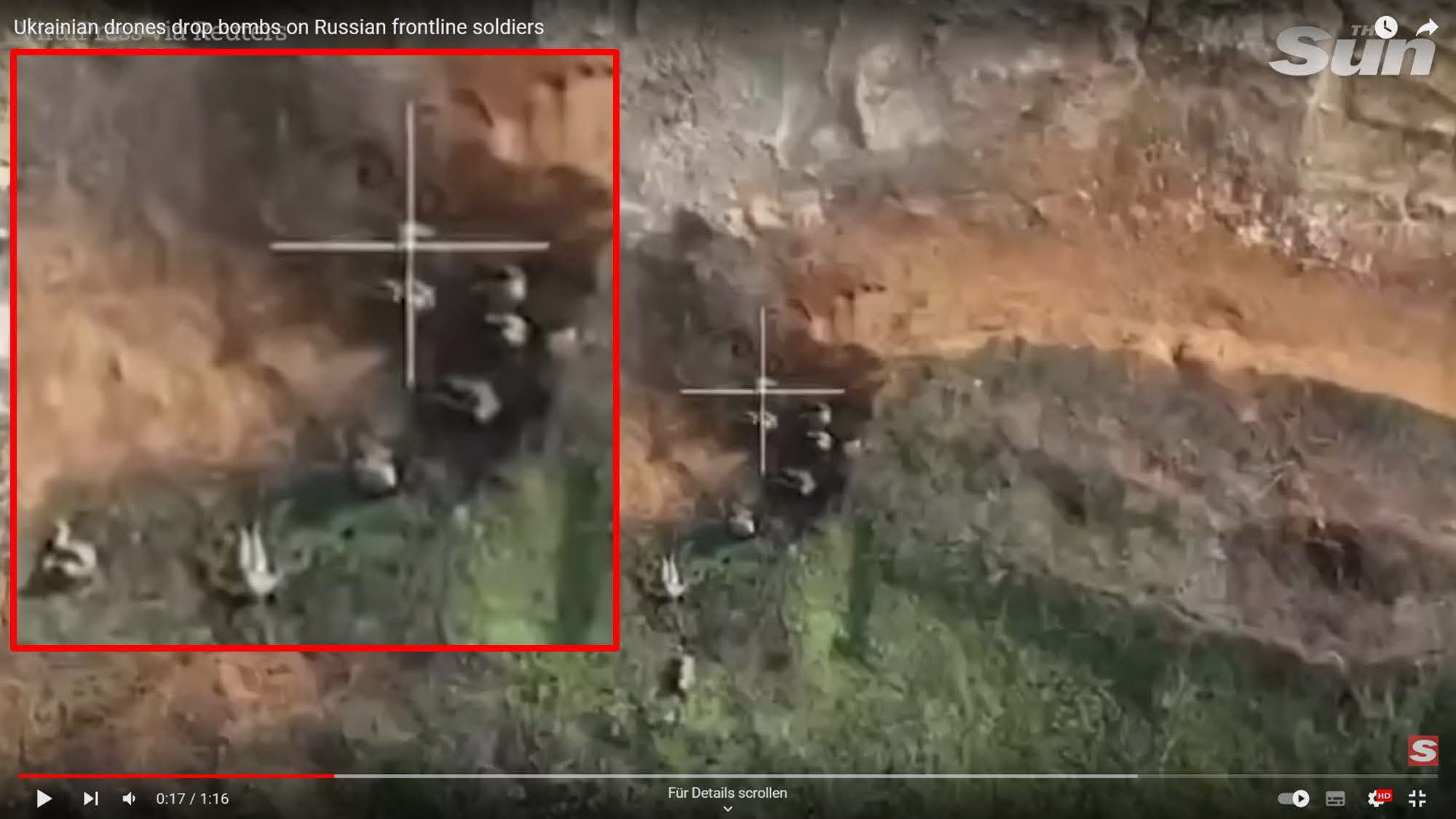}
  \vfill
  \includegraphics[width=0.49\textwidth]{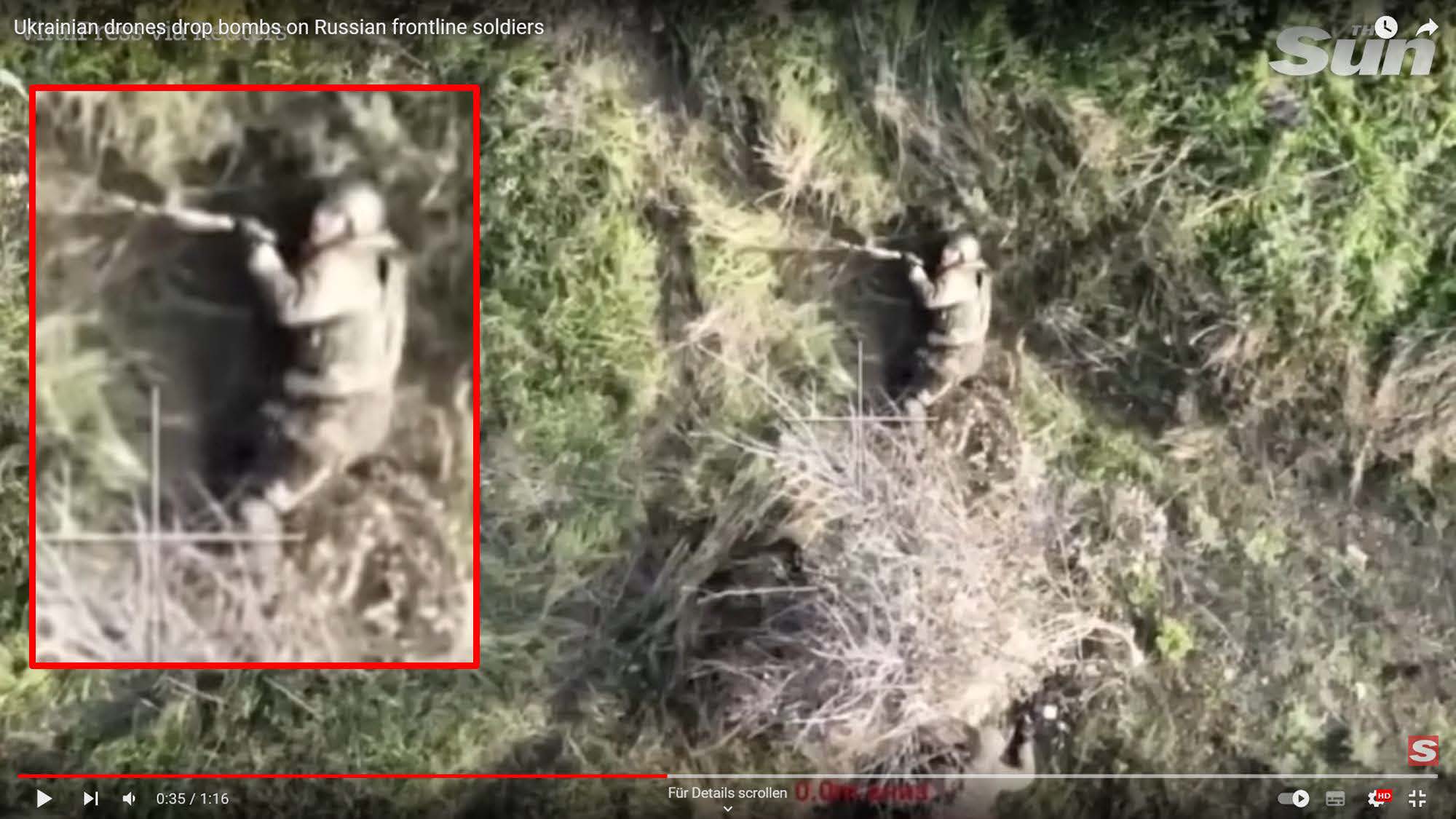}
  \hfill
  \includegraphics[width=0.49\textwidth]{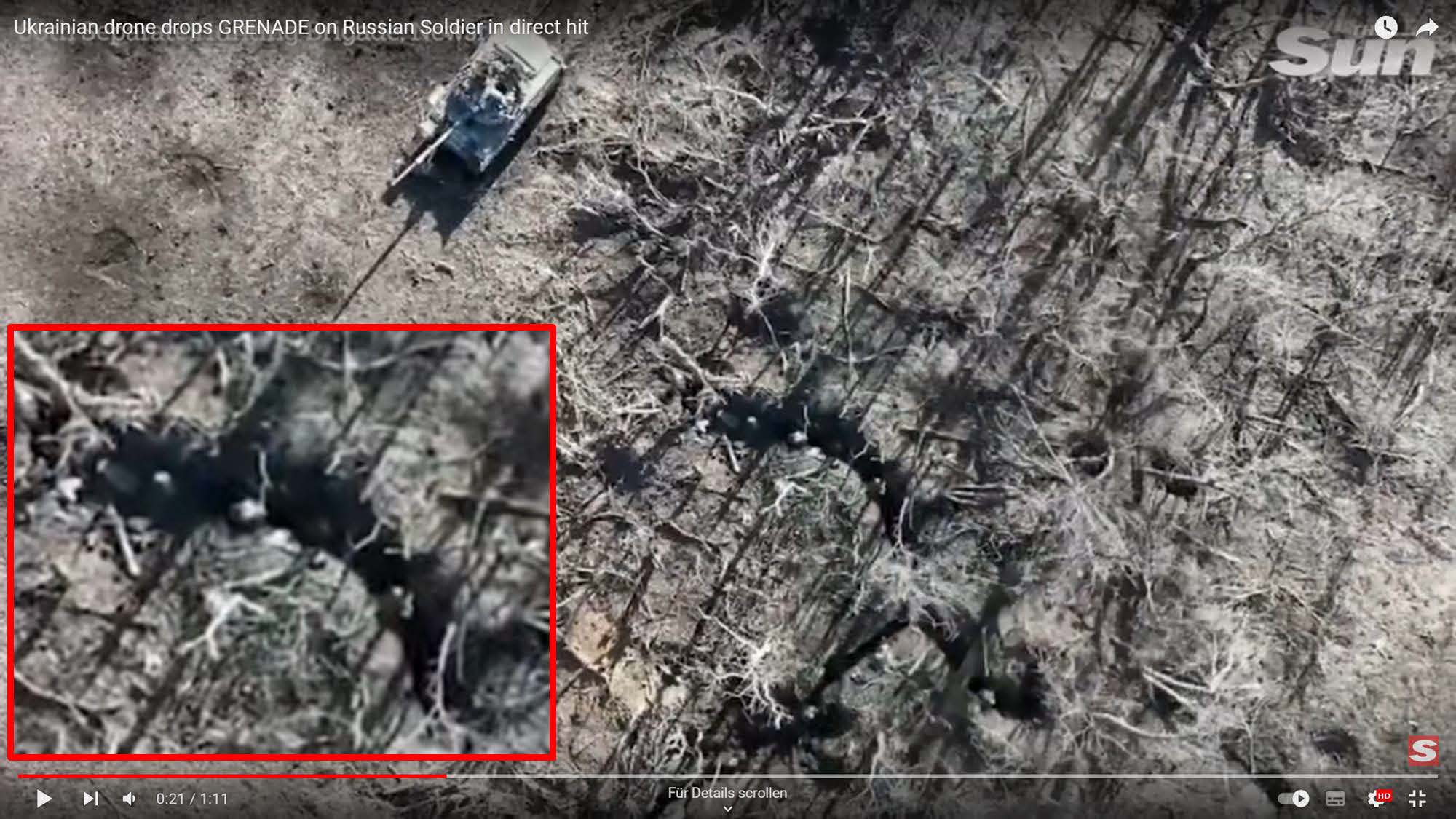}  
  \caption{Images taken from drones of soldiers in the field. Image one to five show wounded or dead persons, image four shows a medical unit with stretcher, six to eight show soldiers hiding. All images are taken from different YouTube videos.}
  \label{fig:drone_images}
\end{figure*}

Detecting and evaluating images or videos on both domains (EO and IR), respective the fusion of both sensor outputs, is a well researched field for the use of artificial intelligence. Based on video recordings, which are available in large numbers on the Internet (see figure \ref{fig:drone_images}), a dataset could be built up, which can be used as a basis to train a machine-learning method for recognising situation-based protective emblems. 
\\\\
In a first step, one has to define all possible passive protective emblems (see section \ref{sec:def_pass_sign}). As mentioned before, these are characterised by wearing the classic protective emblems or evaluation of situations. In a next step, data is to be collected. This can be taken either form the internet (social media, video platforms, etc.) or from internal video documentation. After annotating this data, an AI-model can be trained to perform at least a simple binary classification and inform a downstream routine about the recognition of protective emblems. This information could than be used to either inform an operator and ask for abort or abort automatically and disarm. 
\\
\paragraph{Data collection}
To ensure a wide applicability, one should use data from various platforms, hence EO and IR sensors. The homogeneous distribution of all scenarios which show combatants and non-combatants in different contexts, should also be ensured, as well as the equal share of day and night vision, so that the algorithm can be used 24 hours a day. 
\\\\
It is of great importance to pay attention to a homogeneous distribution of different classes within the dataset, as well as to record a wide spectrum of different situations. This could ensure a great level of abstraction and generalisation capability, that is needed to distinguish various situations. 
\paragraph{AI model and training}
As the routine should run on different platforms, which might be low in computation capacity, a model should be chosen that does need less computational power. Also, this model should be able to deal with the fusion of EO and IR sensors. The recognition of the passive protective emblem s about image classification, a according algorithm is to be taken. 
\\\\
Pose estimation is a technique in machine learning, where the actual state of humans is classified by analysing their posture and sometimes their mimic. This widely researched field could be adapted to a new dimension, where it could be used to classify the state of combatants and non-combatants. Therefore, new postures will be added to the used models in charge that can transfer learned on the data to be collected. 
\\\\
A machine learning model is to be hardened against adversarial attacks, to omit deliberate manipulation of an input image/video feed.

\subsection{Possible use} 
The use of such an algorithm to recognise active or passive protective emblems is the enhanced, modern realisation of the Geneva Convention. In times of (fully) autonomous warfare, where the location of impact is of great distance from the point of launch, operators might not be able to check for non-combatants at the area of effect. Therefore, on the example of an fully autonomous missile that would possibly make use of active protective emblems, such a system could itself recognise the presence of protected facilities and disintegrate on its own. Another example for passive emblems could be the use of autonomous drones. Those still interact with a human operator and could therefore either ask for confirmation, if a protective emblem is present or refuse the use of weapons or the drop of a payload automatically. A human-in-the-loop brings the advantage of a manual control.  uvvvm

\section{Possibilities of allocation} 
\label{sec:POA}

A digital, cross frequency emblem could be centrally defined and used by everyone, as is the current emblem. Using a digital emblem however does bring novel possibilities, as each emblem could be individually issued and revoked. However, it also brings novel challenges as preventing and fighting misuse is even more important.

\subsection{Encoding and Certification of Authenticity}
\label{susbsec:ECA}

While the emblem has been universally adopted since its inception and can be seen as a wide success, it has also always struggled with misuse. This can range from good faith out-of-context usage by civilian entities to deliberate misuse to obfuscate legitimate military targets. With the traditional emblem such misuse can be documented by pictures or verbal accounts. This can then lead to public backlash, as well as limited observance of the protective sign \cite{rolle2009emblem}.

A digital, cross frequency protective emblem comes with additional problems regarding misuse. Documenting cases of misuse is more difficult, as they may not be visible to humans or cannot be photographed. Trust that the digital emblem is only used correctly is therefore more difficult to establish: If there is no way to recognise or punish misuse, why trust the emblem in the first place? The issue can be further highlighted by the use case in autonomous warfare: A drone may decide not to strike a target based on the presence of the emblem. Depending on the, usually limited, communication of the drone, even the party that is operating it may never be notified that a target was avoided due to a emblem. Even if the operators were notified, it might prove difficult for them to establish whether the usage of the protective emblem was justified. As such, there are valid concerns against inclusion of an avoidance of targets marked by a digital emblem. Therefore it is necessary to increase trust in the proper usage of the emblem in order to encourage its introduction, usage and observance. 


\subsubsection{Centralised and Decentralised Systems of Trust}

Problems of trust and misuse are not unique to protective emblems but are ubiquitous throughout the digital domain. Fortunately this means that there are already widely used solutions in place. The most common are public key certificates, also known just as digital certificates, used in authenticating the validity of websites and emails. They are the basis for Transport Layer Security (TLS) which is the basis for the Hypertext Transfer Protocol Secure (HTTPS), a secure encryption standard for web browsing \cite{dierks2008transport}. Digital certificates have to be signed by trusted organisations in order to be valid. This results in a centralised structure as both sides have to trust the same trusted organisation which guarantees the authenticity. Ultimately this results in a hierarchy of trusted institutions known as the chain of trust. In the use case of web browsing, each browser comes with a list of trusted certification authorities and websites each require a certificate which was signed by one of these authorities. Notably it is also possible for the certification authority to revoke a certificate by issuing a signed statement. This process is very important for a protective emblem, since it allows to revoke misused emblems or even revoke all emblems used by an offending faction.

More recently decentralised systems which do not rely on trusted third parties have been developed which are colloquially known as blockchains. A blockchain is an append-only database in which each data package, or block, can be appended to the chain after it was validated that it fits the requirements defined by a common protocol and the previous blocks. Each block contains a cryptographic hash of all previous blocks which makes it very difficult to manipulate the content of blocks. Multiple copies of the chain exist and if a newly added block is deemed invalid they will not be reproduced on the other copies. Mechanisms exist to prevent simultaneous appending by multiple parties. This is known as the double spend problem, as blockchains are most used for digital currencies. These mechanisms rely on solving complex computations ("proof-of-work", POW) or ownership of portions of the digital currency ("proof-of-stake", POS). Overall this leads to a design in which trust is not relegated to a singular entity but rather is guaranteed by a consensus of the largest part of the network, defined as either those with most computational resources for POW or the most tokens in POS \cite{zheng2017overview}. In this use case each protective emblem could be added inside a new block to the chain and be validated by the other users of the same blockchain. This is similar to the current blockchain use case of non-fungible tokens (NFTs). Verification of authenticity is straightforward as only its existence in the chain needs to be checked. Ownership of the token can also be established, which for protective emblems would be the identification of the party fielding the protective emblem.

Both approaches come with a degree of complexity, but blockchain-based systems are notoriously complex to implement and maintain. The consensus needed to coordinate the authorisation and revocation of protective emblems will be difficult to achieve in general. That is a political problem as parties with naturally opposing interests will need to reach a conclusion, with some parties overruling others. This can be achieved in a central organisation such as the UN. A decentralised system of trust does not solve this fundamental issue, instead it tries to represent the rules for consensus algorithmically, which is difficult, complex and error prone \cite{islam2020critical}. The perceived benefit of a decentralised system also does not hold up in practice, as both the development and maintenance i.e. governance of the system often result in a degree of centralisation \cite{sai2021taxonomy}. For this use case, a central, neutral authority is more practical, as it already exists for the traditional emblem in the form of the International Committee of the Red Cross. 

\subsubsection{Discussion of the Feasibility of Digital Authentication}

The standard used for digital certificates by TLS is X.509. A certificate in this format usually requires one to two kilobytes of storage. While this is not a lot for modern communication systems, it does represent a significant amount if it should be represented in a cross-frequency emblem. For comparison, a QR-code can hold two kilobytes of information, passive RFID tags one kilobyte\cite{soon2008qr,chadwick2003role,weinstein2005rfid}. Depending on the method of communication it would therefore be necessary to further compress or reduce the information in the certificate.  

In summary, digital certification of authenticity provides a way to both install and remove trust in a specific emblem, or the party that used the emblem. This is paramount for the acceptance of protective emblems as it enables to punish misuse, which is of even greater importance for digital emblems.   

\section{Ethics} 

With a new protective emblem and the ability of weapon systems to recognise protected individuals, new ethical challenges arise. 

First, with a frequency that can disable long distance weapon systems or at least abort attacks, it would be possible to use this frequency also for areas and buildings which are not officially protected. Using a visual emblem, like the red cross on white, a human would most likely still be able to tell the difference between an actually protected facility and a misuse of the international protective emblem. Using a virtual protective emblem which automatically disables weapon systems or at least halts an attack until a decision was made on whether the facility is actually protected, could potentially invite misuse since the result would be more useful. 

Furthermore, all weapon systems would have to equipped with the sensors and the software necessary to read and react to the protective emblem. This would require all nations to participate in this effort. However, since no software is completely fail proof, disabling the feature and attributing it to a software failure would be possible. 

Similarly, intentional blocking or spamming of signals, for example radar signals, could arise as a new tactic. For the active protection emblem as described above it would be necessary to emit and receive radar signals. Using other senders to intentionally block the radar signals emitted by the protected facility would make it vulnerable to attacks. 

\newpage

\appendix

\bibliographystyle{unsrt}
\bibliography{bib.bib}

\end{document}